\documentclass[traditabstract]{aa}
\usepackage{txfonts}
\usepackage{natbib}
\usepackage{graphicx}
\bibpunct{(}{)}{;}{a}{}{,} 
\usepackage{longtable}
\usepackage{xspace}

\begin{document}

\title{ A mid-term astrometric and photometric study of Trans-Neptunian Object (90482) Orcus}

\author{J.L. Ortiz \inst{1}
\and A. Cikota \inst{2}  \and S. Cikota \inst{2} \and D. Hestroffer\inst{3} \and A. Thirouin \inst{1} \and N. Morales \inst{1} \and R. Duffard \inst{1} \and R. Gil-Hutton \inst{5}  \and
P. Santos-Sanz\inst{1,4} \and I. de la Cueva \inst{6} }


\institute{Instituto de Astrof\'{\i}sica  de Andaluc\'{\i}a - CSIC, Apt
3004, 18008  Granada,  Spain. \and Physik-Institut, Universitat Zurich, Winterthurerstrasse 190, CH-8057, Zurich, Switzerland. \and IMCCE/Observatoire de Paris, CNRS, UPMC, 77 av. Denfert-Rochereau F-75014 Paris, France.
\and Observatoire de Paris, LESIA-UMR CNRS 8109, 5 place
Jules Janssen F-92195 Meudon cedex, France. \and Complejo Astron\'omico El
Leoncito (CASLEO-CONICET) and San Juan National University, Avda. de
Espa\~{n}a 1512 sur, J5402DSP, San Juan, Argentina.
\and Astroimagen. C. Abad y Lasierra, 58 Bis - 62, 07800 Ibiza, Islas
Baleares, Spain.}

\titlerunning{Orcus as test case for binary detection through astrometry}

\abstract{From time series CCD observations of a fixed and large star field that contained the binary trans-Neptunian Object (90482) Orcus (formerly 2004 DW), taken during a period of 33 days, we have been able to derive high-precision relative astrometry and photometry of the Orcus system with respect to background stars. The right ascension residuals of an orbital fit to the astrometric data revealed a periodicity of 9.7 $\pm$  0.3 days, which is what one would expect to be induced by the known Orcus companion (Vanth). The residuals are also correlated with the theoretical positions of the satellite with regard to the primary. We therefore have revealed the presence of Orcus' satellite in our astrometric measurements, although the residuals in declination did not show the expected variations. The oscillation in the residuals is caused by the photocenter motion of the combined Orcus plus satellite system around the barycenter along an orbital revolution of the satellite. The photocenter motion is much larger than the motion of Orcus around the barycenter, and we show here that detecting some binaries through a carefully devised astrometric technique might be feasible with telescopes of moderate size. We discuss the prospects for using the technique to find new binary Transneptunian Objects (TNOs) and to study already known binary systems with uncertain orbital periods. We also analyzed the system's mid-term photometry in order to determine whether the rotation could be tidally locked to the satellite's orbital period. We found that a photometric variability of 9.7 $\pm$ 0.3 days is clear in our data, and is nearly coincident with the orbital period of the satellite. We believe this variability might be induced by the satellite's rotation. In our photometry there is also a slight hint for an additional very small variability in the 10 hr range that was already reported in the literature. This short-term variability would indicate that the primary is not tidally locked and therefore the system would not have reached a double synchronous state. Implications for the basic physical properties of the primary and its satellite are discussed. From angular momentum considerations we suspect that the Orcus satellite might have formed from a rotational fission. This requires that the mass of the satellite would be around 0.09 times that of the primary, close to the value that one derives by using an albedo of 0.12 for the satellite and assuming equal densities for both the primary and secondary.}


\maketitle

%

\section{Introduction}

Trans-Neptunian Objects (TNOs) are important bodies because they are thought to be leftovers from the process of the formation of the solar system and they carry important information about the early stages of the solar system \citep{Morbidelli2005, Tsiganis2005, Gomes2005}. They are also thought to be the parents of the short-period comets \citep{Fernandez1980} and therefore a source of objects that eventually can come close to the Sun or to the Earth.
Among the TNOs, there are dwarf planets whose study is important per se, but also because they provide a wealth of information about the physical processes that take or took place in the trans-Neptunian Belt. Large TNOs are supposed to retain primordial information about the original spin rate distribution because apparently they are the least collisionally evolved objects \citep{DavisFarinella1997, Benavidez2009}. However, some degree of spin evolution owing to tidal interactions in binaries can alter this concept and, Orcus may represent a good example as we will see later, as does Pluto.

The trans-Neptunian object (90482) Orcus (also known as 2004 DW from its provisional designation) is one of the brightest known TNOs discovered so far and possibly one of the largest. Indeed, Orcus qualifies to become a dwarf planet because of its large diameter (D=850$\pm$90 km), which has recently been measured with enough precision by the \textit{Herschel Space Observatory} \cite{Lim2010} and is consistent with \textit{Spitzer} measurements \citep{Stansberry2008}. It belongs to the plutino dynamical class and it is therefore the largest plutino immediately after Pluto. Besides,  Orcus is an interesting object for other reasons: It is known to posses a satellite, Vanth, which orbits Orcus in around 9.5 days  and whose orbital plane is almost perpendicular to the line of sight \citep{Brown2010}. Water ice and perhaps even ammonia has been found on its surface through near infrared spectroscopy \citep{Fornasier2004,Trujillo2005,DeBergh2005,Barucci2008}.

Also, its short term variability was studied in \cite{Ortiz2006} who found a likely rotation period of 10.08 hr (although periods at around 7 hr and 17 hr were also possible). Later, \cite{Thirouin2010} included more data, obtaining a rotation period of 10.47 hr. In both works the variability was very low ($\leq 0.04$ mag). Other works on the short-term variability of Orcus by Sheppard et al. (2007) and \cite{Tegler2005} failed to find a high amplitude periodicity in Orcus, but those works did not reject the possibility of a lightcurve with an amplitude below 0.06 mag (\cite{Sheppard2007}), which is consistent with the Ortiz et al. (2006) results, and the 0.02mag variability in 7 hours of observation reported by Tegler et al. (2005) is particularly consistent with the Ortiz et al. (2006) and Thirouin et al. (2010) rotational lightcurves.

These results already seem to indicate that Orcus' rotation is not tidally locked to its satellite, but because the satellite orbital period is much longer than the usual observing windows for rotational variability studies, the question arises as to whether the Ortiz et al. and Thirouin et al. works could have detected a rotation period as long as 9.5 days.  Therefore we decided to schedule a specific long observing run spanning more than 20 days in order  to study the photometric behavior of Orcus to check whether a long 9.5 day rotation period was possible or not.


We also intended to check whether the presence of Orcus' satellite could be detected by means of high-precision relative astrometry with respect to background stars in order to test the technique for future detection of new binaries by means of telescopes other than the Hubble Space Telescope (HST). Besides, the technique might help in determining the orbital periods of the known binaries whose orbits are very uncertain.
Because Orcus' satellite separation is around 0.3 arcsec, with a small mass ratio, these observations seem challenging, but because Orcus is also among the brightest TNOs, we decided to test the technique with a small telescope (which can easily provide the needed large field of view).

In this paper we report the results from our long astrometry and photometry runs on Orcus. In the first section of the paper we describe the observations and the applied basic image reductions. A second section is devoted to showing the results and their analysis, a discussion section follows and  finally a conclusions section summarizes our main findings.

\section{Observations and reductions}

The CCD images were taken with a 0.45m f/2.8 remotely-controled telescope located on top of Cerro Burek (Complejo Astron\' omico el Leoncito, CASLEO) in Argentina, and equipped with a large format CCD camera of  4008 x 2672 pixels. The pixel scale is 1.47 arcsec/pixel and the total FOV of the instrument is 98$\times$65 arcmin. The observations were obtained through a very broad-band filter in order to maximize the signal-to-noise ratio. The transmission curve is shown in Fig. 1. Integration times were always 300 s and the telescope was always tracked at sidereal rate. The trailing of the object during these short times was negligible. The observations were taken during 18 nights spanning a period of 33 days. A total of 180 images were acquired for this project. The typical signal-to-noise of the Orcus' observations was around 30. The images were usually taken near the meridian so that the object was at its highest elevation as seen from Cerro Burek; this maximizes the signal-to-noise ratio that can be achieved and at the same time minimizes the differential refraction. Seeing ranged from 2 to 4 arcsec, and therefore the Orcus-Vanth pair was always unresolved.

In each observing night we aimed the telescope at fixed coordinates so that the observed star field was exactly the same at all dates of observations. In other words, the images were not centered on Orcus. A key issue in our program was that the field of view of the instrument is very large, which allowed us to always use the same reference stars for the photometry (and that is also true for the astrometry). Therefore we could perform very high precision relative photometry and relative astrometry. In other words, our project could be carried out because it was executed with a large FOV instrument. This would not have been possible with the much smaller field of view of most large telescopes.

\begin{figure}
\centering
\includegraphics[width=8cm]{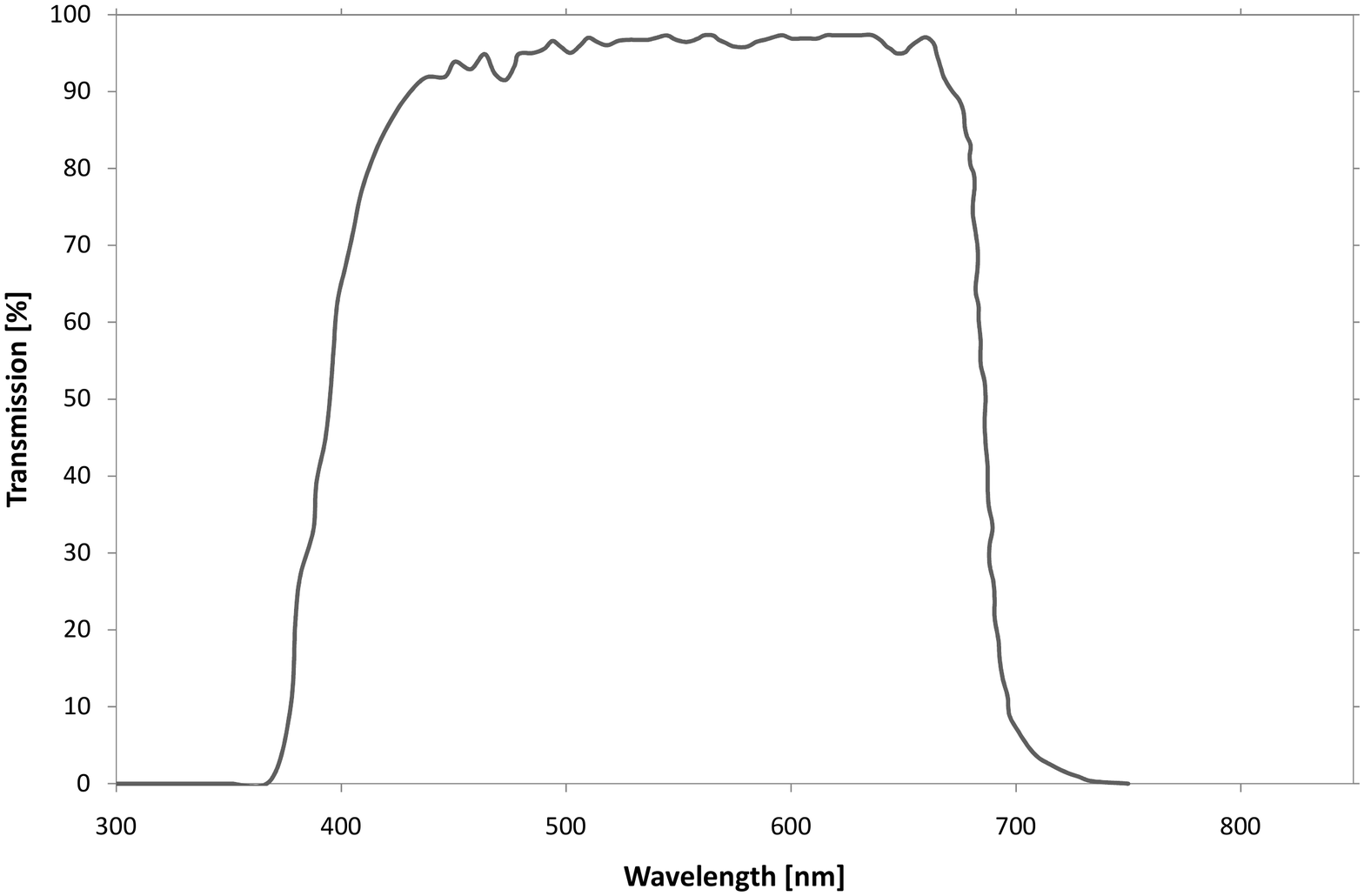}
\caption{Transmission curve of the filter used in this work.}
\label{fig1}
\end{figure}

The images were corrected for bias and dark current by means of  master bias and dark current frames obtained by median combining 10 to 20 images on average. Flatfield corrections were also applied with median flatfields taken at dusk. An image of the observed field with the motion of Orcus indicated is presented in Fig. 2.

The astrometry was obtained by finding the best third-order polynomial that related the image coordinates and sky coordinates. In order to solve the equations we used $\sim$ 500 UCAC2 reference stars. The UCAC2 astrometric catalog \citep{Zacharias2004} was used because it offered a convenient number of reference stars in order to solve the plates. However, the choice of the catalog was irrelevant because our goal was to obtain high accuracy relative astrometry, not absolute astrometry. The choice of any other catalog would be acceptable as well, as long as the catalog has enough stars to reliably solve for the polynomial plate constants.  The source positions were derived by using SExtractor \citep{Bertin1996}. The aperture radius for finding the centroids of the Orcus-Vanth system was 3 pixels. Because the image scale of the detector is 1.47 arcsec/pixel, the 3-pixel aperture guaranteed that most of the flux of the objects entered the aperture even for the poorest seeing conditions possible. The typical uncertainties in the astrometry were slightly larger than a tenth of the pixel size. An average uncertainty of 0.13 arcsec was determined from the measured and known positions of the UCAC2 standars. Nevertheless, because Orcus is fainter than the UCAC2 stars and its centroid determination would be more noisy, we measured the standard deviation of the positions determined for stars of similar brightness to Orcus. The standard deviation turned out to be 0.18 arcsec. Note that these are uncertainties of the individual images. By using large numbers of images one can pinpoint motions smaller than 0.18 arcsec.


The relative photometry was obtained by following a similar approach to that in Ortiz et al. (2006), although with some improvements as described in \cite{Thirouin2010}. The positions of 20 reference stars used for the relative photometry are also marked in Fig. 2. Several synthetic aperture radii were used, but we chose the aperture that resulted in the best photometry in terms of scatter. Images that showed Orcus to be close to a faint star or had potential problems for the high-precision photometry were discarded. Nearly 20\% of the images were rejected for use in the relative photometry analysis. Some stars that showed variability were rejected from the analysis as well. The final standard deviation of the photometry was 0.07 mag. Because Orcus' phase angle and distance to Earth and Sun changed somewhat along the 33-day period, the data were corrected for these effects by subtracting a $-5log(r \Delta)$ factor (where $r$ is distance to the Sun and $\Delta$ is distance to Earth). Those distance-corrected relative magnitudes were used to fit a linear phase dependence, which is known to work well for TNOs (\cite{SheppardJewitt2003}). The phase slope we derived is 0.08 $\pm$ 0.04 mag/degree, which is consistent with the value of 0.11 $\pm$ 0.03 from \cite{Rabinowitz2007} in V band. After this phase dependence was removed we obtained magnitudes as a function of date, which were later used to determine a lightcurve as described in the next paragraph.


\section{Results and analysis}

Tables 1 and 2 contain the final astrometry and the relative photometry (corrected for distance and solar phase angle) respectively.

\begin{figure}
\centering
\addtocounter{figure}{+1}
\includegraphics[width=7cm, angle=90]{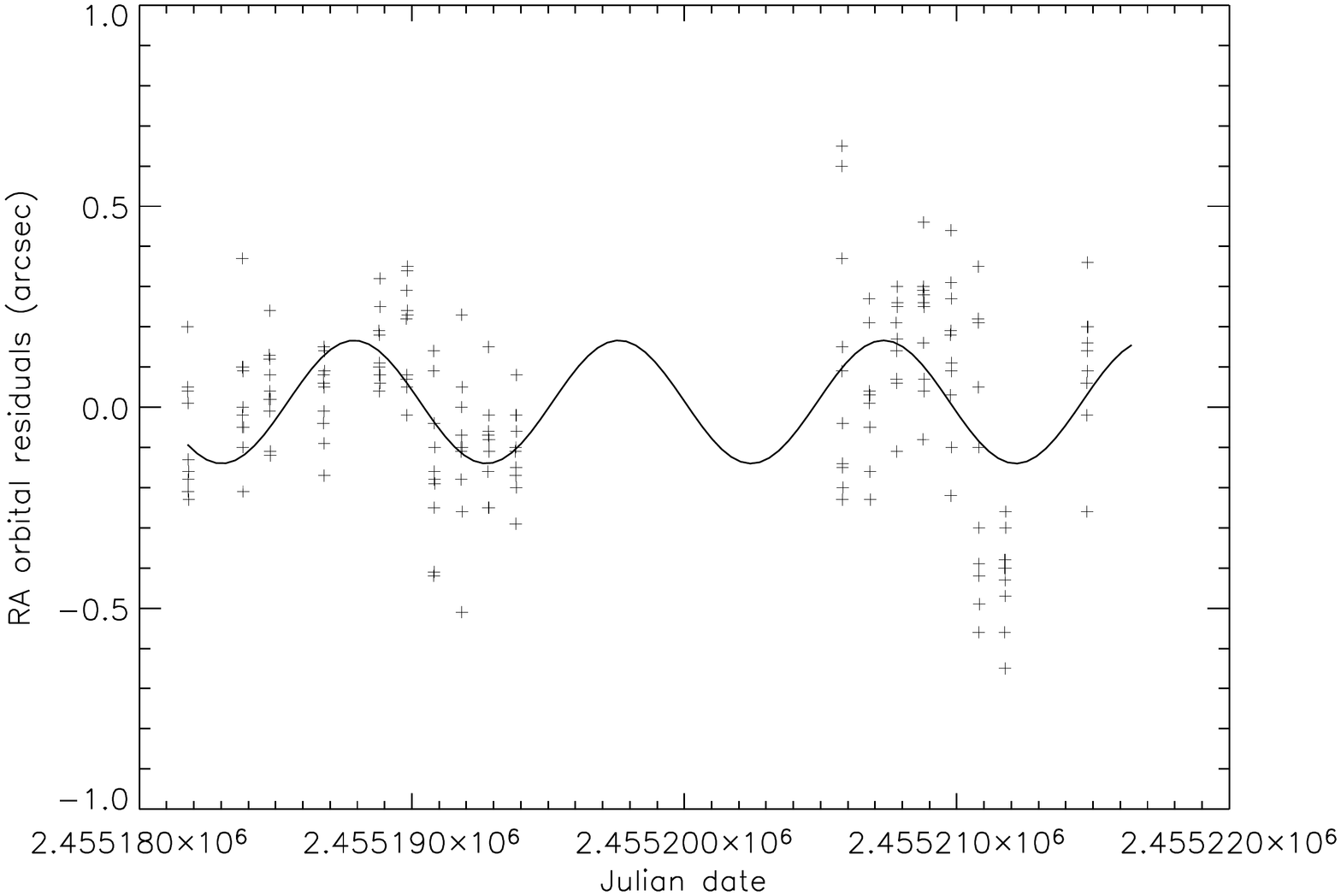}
\caption{Right ascension residuals as a function of date from an orbital fit to the astrometry in Table 1. A sinusoidal fit to the data is superimposed.}
\label{fig3}
\end{figure}

Concerning the astrometry results, the right ascension (RA) residuals obtained from an orbital fit to the astrometry are shown in Fig. 3. as a function of date.  A Lomb periodogram analysis \citep{Lomb1976} of the time-series RA residuals is shown in Fig. 4. As can be seen in the plot, the highest peak in the periodogram is at 0.1029 cycles/day, which corresponds to a period of 9.7 $\pm$ 0.3 days. The confidence level of the detection is well above 99\%. Such a  period is entirely consistent with the 9.53-day orbital period of Orcus's satellite \citep{Brown2010}. >From a sinusoidal fit, the peak to peak amplitude of the oscillation in the residuals is  0.3 $\pm$ 0.2 arcsec.

If most of the orbits of binary systems lie on the ecliptic, we expect that the RA residuals are more appropriate than the declination residuals to study the systems because the declination residuals would be more difficult to detect in these cases. However, because Vanth's orbit plane appears to be close to the perpendicular to the line of sight, the residuals in declination should also reveal the periodicity. However, we did not find the 9.7-day period. There are several reasons that can explain this. They are discussed in the next section.

\begin{figure}
\centering
\includegraphics[width=7cm, angle=180]{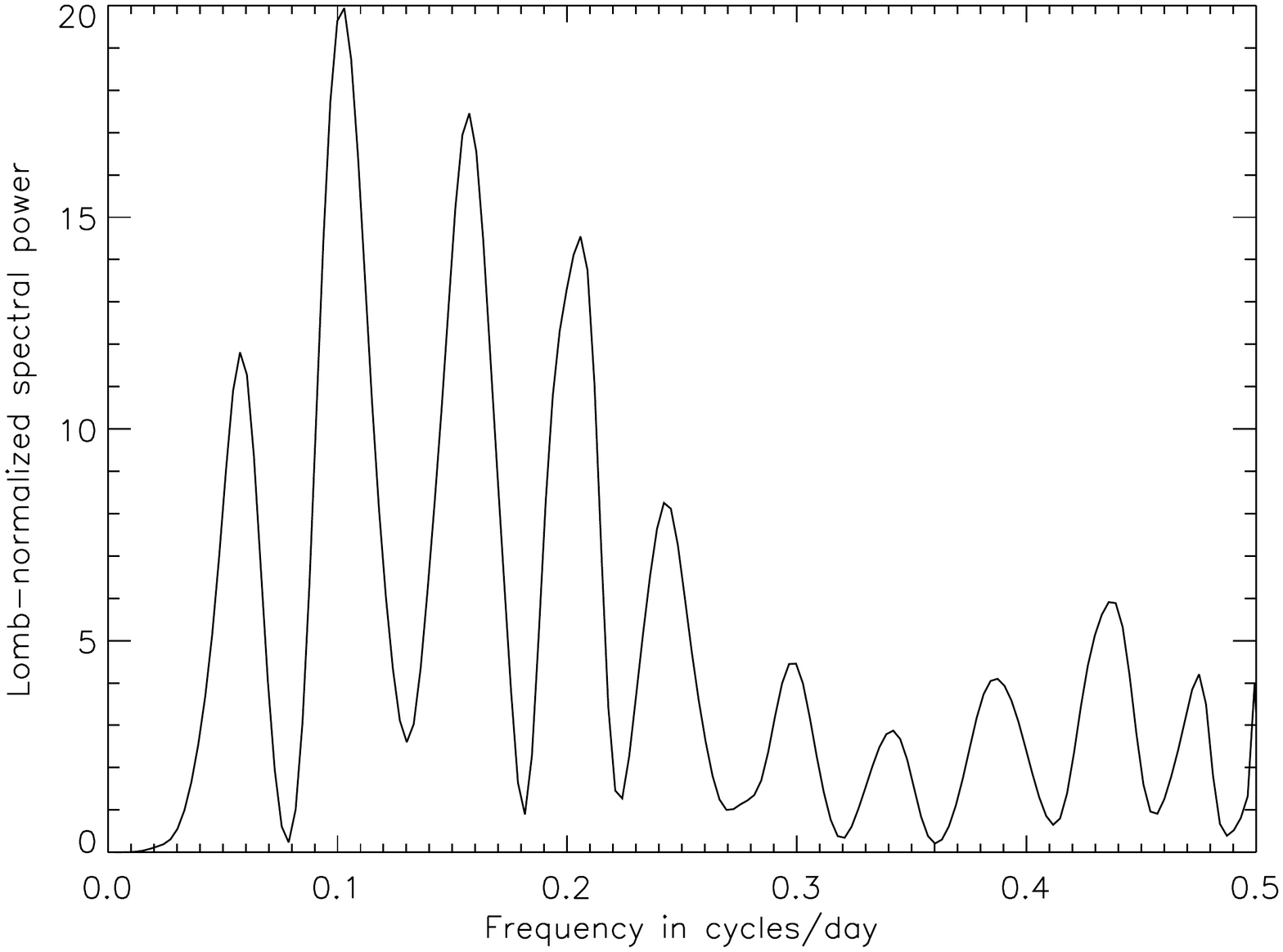}
\caption{Lomb periodogram of the right ascension residuals. The spectral power is plotted as a function of frequency (in cyles/day).}
\label{fig4}
\end{figure}

We also studied whether the values of the residuals were correlated or not with computed theoretical positions of Orcus' satellite. We did that as a further test to check whether we had indeed detected the presence of a satellite in our data or if the result was a mere coincidence (despite the very high significance level of the detected periodicity). We took nightly averages of the residuals to avoid computing around 200 orbital positions. The binned residuals in arcsec and the theoretical east-west distance of the satellite with respect to Orcus  are shown in Table 3. The theoretical positions were computed with the orbital information given in \cite{Brown2010} and updated in Carry et al. (2010, submitted). A Spearman test results in a clear correlation of the two columns in Table 3 with a significance level of 97$\%$. We used the Spearman test because this correlation analysis is independent of the exact functional form of the relation, which is not known a priori. Although the angular separation of the satellite with respect to the primary should be linearly related to the theoretical distance between primary and secondary, the photocenter separation in groundbased observations is a complex function of the expected angular separation, seeing, observing conditions, and magnitude difference of the primary to the satellite. Nevertheless we have also performed a linear regression analysis, and the corresponding fit is shown in Fig. 5. The coefficients of the fit were 0.003 $\pm$ 0.030 for the intercept and 1.39$\times$10$^5$ $\pm$ 0.49$\times$10$^5$ for the slope. The periodogram and the correlation analysis are two different diagnostics, and which indicate the presence of astrometry residuals linked to the satellite. We can thus be confident that the presence of Orcus' satellite is unambiguously revealed in our data.

\begin{figure}
\centering
\includegraphics[width=7cm, angle=90]{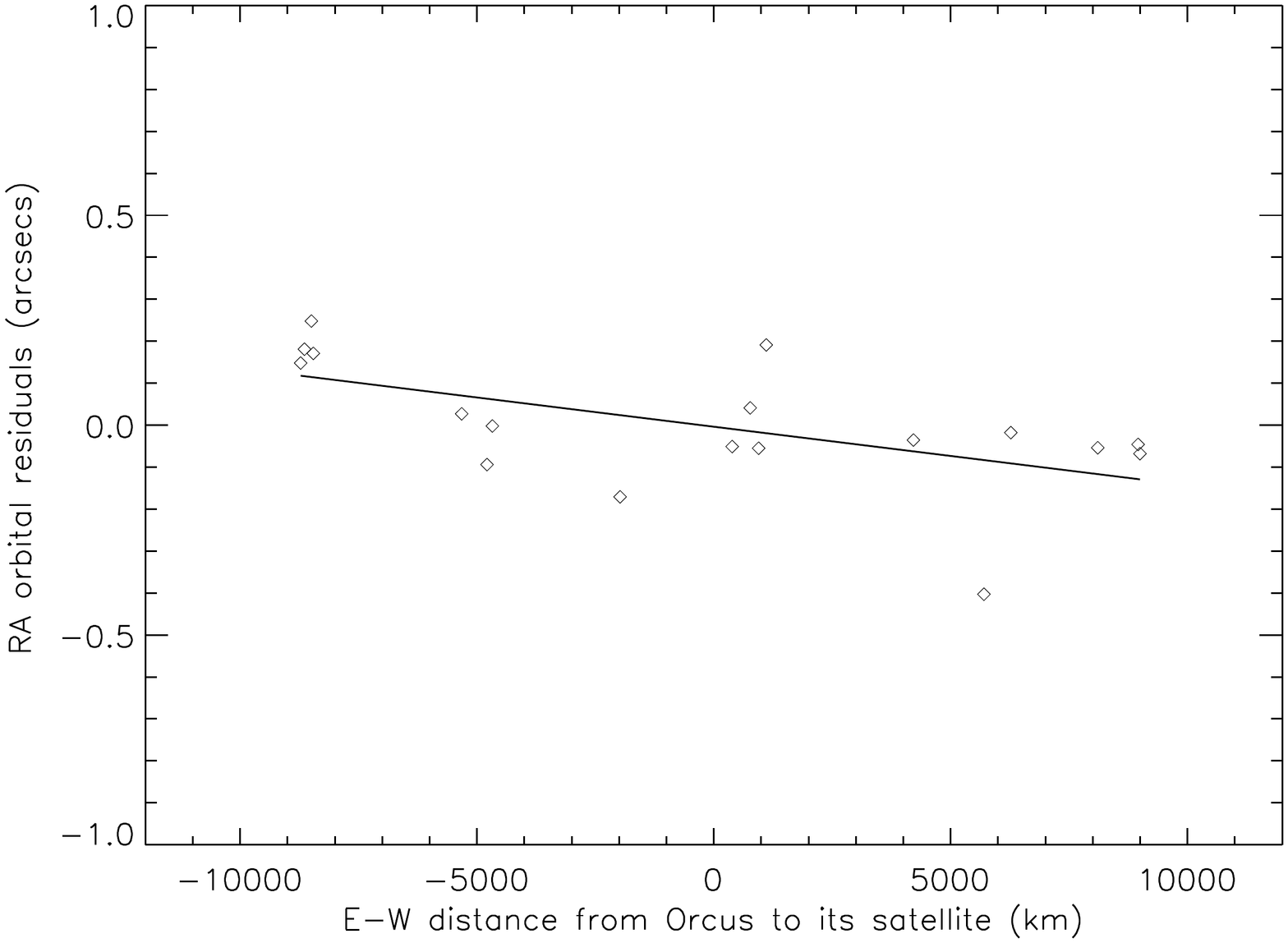}
\caption{Linear fit to the residuals versus computed E-W distance of the secondary to the primary (distance to the East is taken as negative).}
\label{fig5}
\end{figure}

On the other hand, the periodogram analysis of the time-series relative photometry clearly indicated a frequency of 0.1029 cycles/day, which corresponds to a period of 9.7 $\pm$ 0.3 days (see Fig. 6). It must be pointed out that the 9.7-day variability was already detected even prior to correcting the data for solar phase angle and distance to Earth and Sun. Hence, it is not an artifact from the data processing. A lightcurve for that rotation period is presented in Fig. 7. The peak to peak amplitude from a sinusoidal fit to the data is 0.06$\pm$0.04 mag. This variability could becaused by a nonspherical shape or albedo variations or even a combination of both. In addition, the 9.7 $\pm$ 0.3 day variability could be caused by the primary or by the satellite. We suspect it is the satellite, for reasons that will be discussed in the next section.

\begin{figure}
\centering
\includegraphics[width=7cm, angle=180]{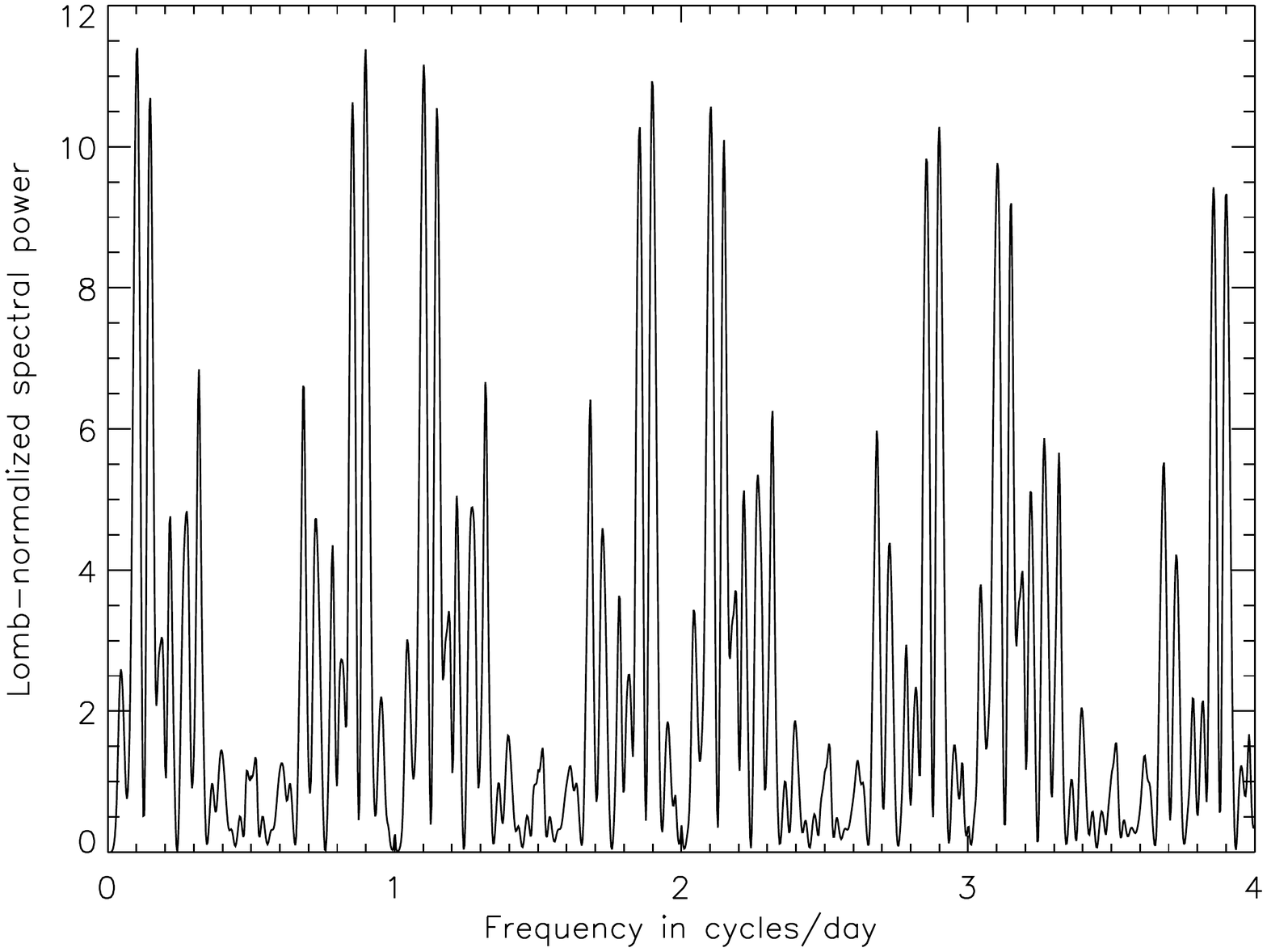}
\caption{Lomb periodogram of Orcus' relative photometry. The spectral power is plotted as a function of frequency (in cycles/day) because this is a convenient way to easily identify possible aliases of the main frequency. See text.}
\label{fig6}
\end{figure}

\section{Discussion}

The predicted position for Orcus based on its orbit around the Sun should basically correspond to the barycenter of the system, not exactly to that of the largest component of the system. With a nominal mass ratio supposedly of $\sim$ 0.03 (Brown et al. 2010), the offset (primary to center-of-mass) could be $\sim$250km in distance. At Orcus' distance from Earth, and neglecting the light contribution of the secondary, this translates into a mere $\pm$0.009arcsec wobble, which would be undetectable in our data.
Therefore, it appears that the light contribution of the secondary must be very relevant.

The maximum separation of Orcus and its satellite is around 9000 km. At Orcus's distance from Earth this translates into approximately 0.3 arcsec. Because the brightness of Orcus' satellite is not negligible, it might shift the photocenter a large enough amount to be detected. Then, the motion of the photocenter around the barycenter (which is very close to the primary) might seem the correct explanation of the periodic signal that we are detecting in our astrometry. We have modeled the maximum photocenter shift of the combined Orcus + satellite system with respect to the primary by generating synthetic images in which there are two point sources with 0.3 arcsec separation and a magnitude difference of 2.5mag \citep{Brown2010}. These point sources were convolved with Moffat point spread functions (which are typical of ground based observations) for several seeing values, and the position of the photocenter was measured with respect to the position of the primary. The DAOPHOT centroid algorithm was used to find the photocenter. For the typical seeing conditions of our observations the maximum separation of the photocenter with respect to the position of the primary is 0.03 arcsec according to our simulations. Therefore, the peak to peak variation in the residuals of our astrometric observations should be around 0.06 arcsec, which is much larger than the barycenter wobble mentioned in the first paragraph, but 0.06 arcsec is less than the 0.3 $\pm$ 0.2 arcsec amplitude of the astrometry residuals that we have measured.

The main parameter to increase the photocenter shift of the simulations to reach the almost 0.3 $\pm$ 0.2 arcsec amplitude in the residuals is the magnitude difference between Orcus and its companion. By reducing it to just 0.5mag we would obtain a nearly satisfactory agreement. However, Vanth's brightness would have to oscillate by nearly 2 magnitudes in a rotation period, which is not feasible: the satellite would have to be too elongated. It appears more likely that the true oscillation in the residuals is closer to the lower end of our estimate (0.1 arcsec), which is compatible with the error bar. From the synthetic images, in order to reach 0.1 arcsec amplitude in the residuals, the magnitude difference of secondary to primary should only change from 2.5mag to 2.0mag. This brightness change in the satellite would induce a 0.06mag lightcurve amplitude on the Orcus system. This coincides with the 0.06 $\pm$ 0.04 mag lightcurve amplitude that we presented here, and therefore the satellite variability might explain both the amplitude of the astrometry residuals and the amplitude of the lightcurve. However, keeping in mind that the orbital plane of the satellite is almost perpendicular to the line of sight, the satellite's spin axis orientation should not be very far from the perpendicular of the orbital plane and in order for a 0.5 mag change to take place with this orientation, the satellite would have to be considerably elongated. A large magnitude change in the satellite' brightness caused by albedo variegations is also a possibility, but high variations are only known for a few objects in the solar system. The saturnian satellite Iapetus, whose leading side is almost 2 magnitudes fainter than its trailing side, is the most extreme case. However, for Orcus it is difficult to envision a similar scenario to that proposed for the existence of Iapetus' two distinct sides. If the real peak to peak amplitude of the RA residuals is even smaller than 0.1 arcsec, then the needed brightness variation of Vanth is smaller than 0.5mag, which would mean that the satellite does not have to be very elongated or present very high albedo variations.



The variability in Vanth can also offer an explanation for the lack of detection of the 9.5 day periodicity in the declination residuals. Because Vath's brightness maxima are nearly in phase with the maxima in RA residuals, the RA residuals are the ones that reach the highest amplitude according to the simulations with synthetic images because the separation is sensitive to the magnitude difference. Other reasons for the lack of detection of 9.5 day periodicity in the declination residuals might be a smaller inclination of the orbital plane than the perpendicular to the line of sight. This might be enough to reduce the amplitude of the residuals so that detectable levels are not reached, or maybe there were systematic effects in declination (like contamination from background stars as Orcus moves with respect to the star field).

>From the Orcus experience we can try to draw some conclusions for the prospects of detecting new binaries by means of the astrometric technique and also for the study of known binaries that have very uncertain orbital periods. Because most of the TNO binary discoveries have been made by means of the Hubble Space Telescope or by means of adaptive optic instruments on large telescopes,for which observing time is scarce, a different approach to detect and study binary TNOs that would make use of other more accessible astronomical facilities might boost this important area of TNO science. From the Orcus experience we have  detected the satellite with a precision in the relative astrometry measurements of around 0.15 arcsec for the individual exposures. This precision can be considerably reduced with larger telescopes. The main cause for the uncertainties in the relative astrometry is the uncertainty in the centroid calculation, which is basically a function of the achieved signal-to-noise ratio and the pixel scale. Therefore, telescopes in the 2m-range should be capable of delivering good signal to noise ratios on m$_v\sim$21 objects and would allow us to detect oscillations in the astrometry of only a few tens of mas. This would inturn allow us to detect close faint companions, even closer than the Orcus satellite. Short orbital periods would be the easiest to detect, because mid to long orbital periods would require long observing runs and very large fields of view. Therefore the technique has the potential to reveal closer binaries than those that HST and adaptive optics systems are finding.


Concerning the photometry results, the nature of the $\sim$10 hr short-term variability of low amplitude reported in Ortiz et al. (2006) and Thirouin et al. (2010) (with very high significance levels) is difficult to asses. We investigated whether this variability is also seen in the present data, but neither the precision nor the observing windows were appropriate to study a very low amplitude short term variability of around 10 hr. However, in the periodogram of Fig. 5 there is a high peak in the 2.3 cycles/day range (corresponding to a period of around 10.4 hr) that is not an alias of the main period (most of the large peaks in the periodogram are aliases of the 9.7-day period). The periodogram in Fig. 5 is shown in the frequency domain because aliases are easily identified as $k+f_0$ or $k-f_0$, where k is an integer and $f_0$ is the main frequency. Thus, the periodogram shows a hint for a possible short-term variability consistent with that reported in \cite{Thirouin2010} and \cite{Ortiz2006} . Unfortunately, the spectral power of the peak is not very high and the significance level of the peak is below 80\%. Therefore, from the present data there is only a hint for the $ \sim$ 10 hr period.



If the short-term variability in \cite{Ortiz2006} and \cite{Thirouin2010} is indeed not an artifact, it might be revealing the primary's spin period (or half its spin period if the variability is caused by shape), which could be different to that of the satellite. A variability of the amplitude $\leq$ 0.04 mag in the primary is below the average 0.1 mag amplitude of the large TNOs whose short term variability has been studied \citep{Thirouin2010, Duffard2009}, but the largest TNOs have even smaller variability than the average because there is a well identified size-dependence \citep{Trilling-Bernstein2006, Sheppard2008, Thirouin2010}. If the spin axis orientation of Orcus is not far from the perpendicular to the satellite orbital plane, the aspect angle would be small and therefore we would expect a smaller variability than the mean 0.1 mag. From the studies of \cite{Duffard2009} and \cite{Thirouin2010} the object is likely to be a McLaurin spheroid with variability induced by albedo variations. The $\sim$ 10 hr rotation rate of Orcus is slower than the TNO average. Assuming that the current rotation rate was primordial and assuming hydrostatic equilibrium (which is likely for TNO this large) the equilibrium figure for plausible densities is a McLaurin spheroid. In summary, the variability in the primary is presumably from albedo variegations, and the $\sim$ 10 hr variability with $\sim$ 0.04 mag amplitude seems entirely consistent with what we know about Orcus. If the primordial spin rate of Orcus were much faster (this will be dealt with at the end of the discussion), one might expect a Jacobi shape and some elongation, but a body like this viewed from the presumed aspect angle near the line of sight would present very small variability. This might well be the case if the object is the remnant of a rotational fission, as discussed in the last paragraph of this section.

One might also wonder whether the photometry might be indicating a rotation plus a forced precession of the primary, with no contribution from the satellite. For single bodies as large as the TNOs, excited rotation states are extremely unlikely because of the very small damping time compared with the age of the solar system (e.g. \cite{Harris1994}. However, for binary bodies one might think that forced precession might be present. Unfortunately our time series data do not cover a time span sufficiently long in order to be able to fit two-dimensional Fourier series like those used to study tumbling asteroids \citep{Pravec2005}.

Therefore we believe that the current best explanation to the available photometry data is that Orcus' satellite has a tidally locked rotation that Orcus has not.

The time required for tidal locking of a satellite around a planet is \citep{Gladman1996, Peale1977}

$$ t = {\omega a^6 I Q \over  3 G m_p^2 k_2 R^5 }, \eqno(1)$$

where
$\omega$ is the initial spin rate (radians per second),
$a$ is the semi-major axis of the orbit of the satellite around the planet,
$I$ is the moment of inertia of the satellite,
$Q$ is the dissipation function of the satellite,
$G$ is the gravitational constant,
$m_p$ is the mass of the planet,
$k_2$ is the tidal Love number of the satellite,
and $R$ is the radius of the satellite.
$k_2$ can be related to the rigidity of the body and its density by means of the equation \citep{Murray2000}

$$ k_2 = { 1.5 \over 1 + 19\mu /( 2 \rho g R)  }, \eqno(2) $$

where $\mu$ is the rigidity of the body, $\rho$ its density and $g$ is the gravity acceleration at the surface of the satellite. Using plausible values of all the parameters, the timescale for the satellite spin locking is much shorter than the age of the solar system. Thus it is likely that the satellite has reached a synchronous state.

The time for the tidal locking of the primary can also be obtained from Eq.1 by swapping the satellite and planet parameters.
That Orcus could still be spinning relatively fast (it would have slowed down only from $\sim$7 hr, the average initial spin in the Kuiper belt \citep{Duffard2009} to 10 hr) would indicate that the mass ratio as well as the diameter ratio of the system is very low, but the Love number and the dissipation factor might also be very different in the primary compared to the satellite.

If we assume that the Orcus primary has been tidally despun from 7 hr to 10 hr we can compute the total angular momentum, lost by the primary, which would have been gained by the secondary in the form of orbital angular momentum and therefore the orbit semimajor axis would have expanded. The estimated initial semimajor axis would be around 3700 km for a mass ratio of 0.03, and about 8300 km for a mass ratio of 0.3.

Regardless of the formation mechanism of the satellites in the solar system, most of the large satellites orbiting planets are tidally locked. Therefore, if the Orcus' satellite is tidally locked, this does not give information on its formation scenario. However, if one assumes that it formed after an impact and the orbit evolved tidally, one can constrain the initial semimajor axis of the satellite orbit assuming a range of masses for the satellite. The values mentioned in the previous paragraph would correspond to the initial configurations after collision.
Alternative scenarios to the collision for the origin of Orcus' satellite might be more appropriate, like the capture mechanism. The semimajor axes determined above would then be the initial configuration after the capture.

If Orcus' primordial rotation rate was already around 10 hr when it formed, then it appears that the tidal interaction of the satellite has not slowed down Orcus significantly. This would either mean that the mass ratio of Vanth relative to Orcus is extremely small or that the formation of the system is relatively recent. Yet the 0.03 mass ratio that we used as the nominal value is already a low value, because it comes from the assumption that the albedos of both Orcus and Vanth are equal, but it is very likely that the albedo of Vanth is much smaller than that of Orcus. Orcus has prominent water ice absoption features in its spectrum, which means that that the surface content of water ice is large and the geometric albedo should be high. That is consistent with the $\sim$ 0.3 albedo determined by several authors. On the other hand the satellite shows no water ice features in its spectrum (Brown et al. 2010) and most likely an albedo of 0.12, which is close to the average value for non-water ice rich TNOs, would be applicable to Vanth. With this value, the mass ratio would approximately be 0.09.

It is suspicious that the spin rate of Orcus is still high despite the tidal interaction.  One may argue that the initial rotation state of Orcus was much faster than it is now. Indeed, if the initial spin rate of Orcus were close to its critical rotation, Orcus might have broken up and the satellite might have been a result of such proccess. The specific angular momentum that Orcus would have had would have been close to the one we observe today in the system, provided that the mass of the satellite is higher than the nominal value of 0.03 times Orcus' mass. Indeed, a value of 0.09 which we derived above for the mass ratio, would provide exactly the needed specific angular momentum. In other words, 0.09 is coincident with the mass ratio obtained by using a lower and more realistic albedo value for Vanth than that of Orcus. When one uses that mass ratio and a 10 hr period for Orcus, the specific angular momentum of the system exactly matches that of a very fast rotating Orcus near its critical limit.
The angular momentum from the slowing down of Orcus' rotation rate would have been transferred to the satellite orbital angular momentum. This transfer would have caused the satellite orbit to move to its current position (from $a$=0 to $a$=8980 km). Therefore we believe that the rotational fission of Orcus is a good candidate mechanism to have formed Vanth. Rotational fission of TNOs is discussed in more detail in Ortiz et al. (2010, in preparation).


\begin{figure}
\centering
\includegraphics[width=7cm, angle=90]{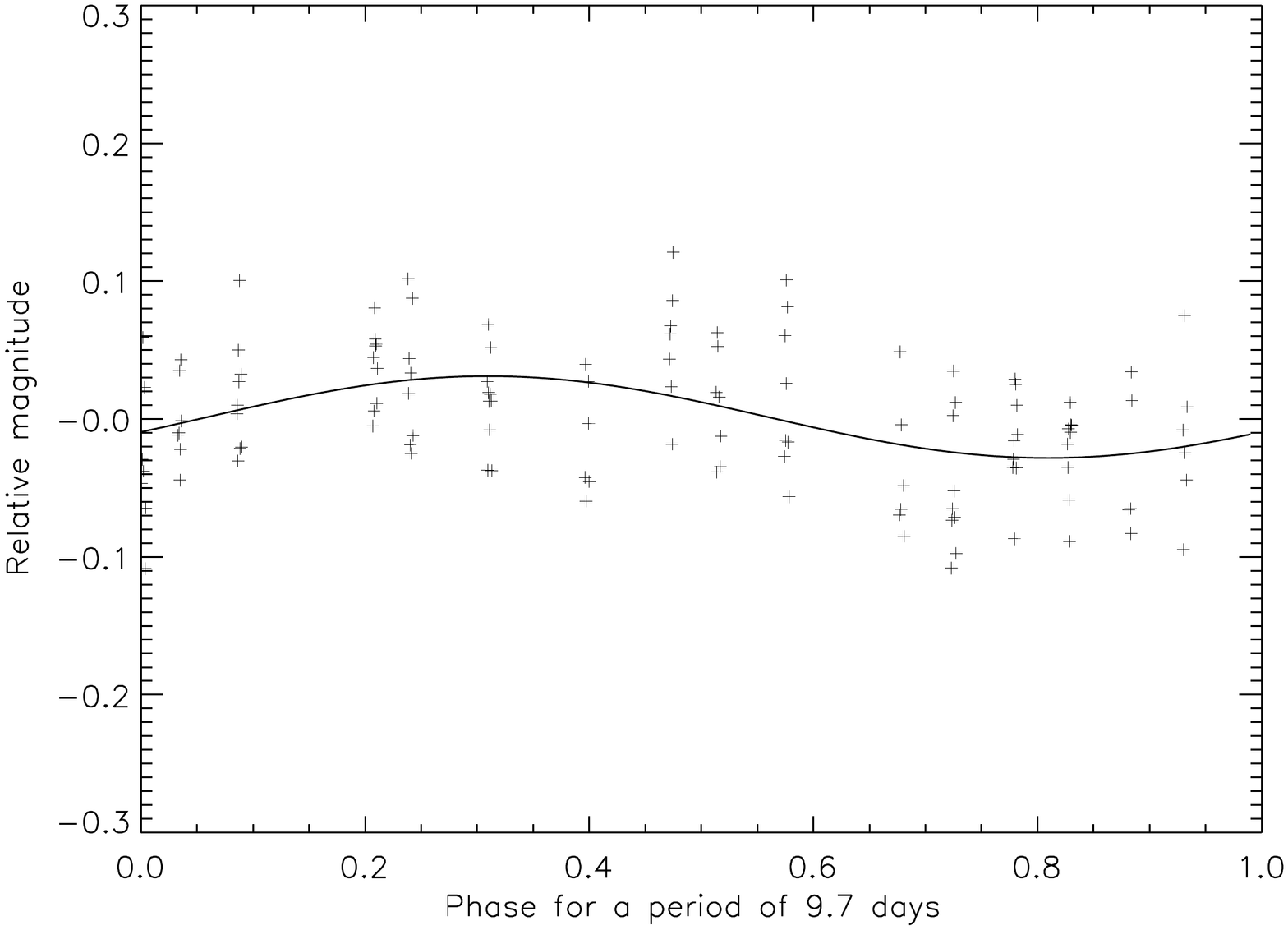}
\caption{Lightcurve resulting from the relative photometry. The relative magnitude is plotted versus the rotational phase (normalized to 1 rather than 2$\pi$). A sinusoidal fit is superimposed.}
\label{fig7n}
\end{figure}



\section{Conclusions}

We presented results from an 18-night astrometry and photometry run devoted to Orcus' system. The results clearly show that Orcus' satellite imprints an unambiguous periodic signal in the relative astrometry, which is detectable despite the high magnitude difference between Orcus and its satellite ($\sim$ 2.5 mag). The periodicity in the astrometry residuals is coincident with the orbital period. The values of the residuals are correlated with the theoretical positions of the satellite with respect to the primary. We have thus shown that detecting binary systems in the trans-Neptunian Belt by means of high-precision astrometry with medium to large telescopes is feasible provided that the barycenter and photocenter of the binary systems do not coincide and are separated by at least tens of milliarcseconds. Because the typical magnitude difference of the binary components is small in the known binaries \citep{Noll2008}, much smaller than in the test case of Orcus, while on the other hand separations of thousands of km are typical among the binary TNOs, the wobble of the photocenter might be detectable. Therefore, specific relative astrometry campaigns with moderately sized telescopes might be a powerful means to study TNOs. Another possible observing strategy is to perform absolute astrometry; this necessitates good astrometric catalogs with faint stars like the astrometric catalog that the Gaia mission will provide. From our photometry run we also determined that Orcus' system has a 0.06$\pm$0.04 mag variability with a period of 9.7 $\pm$ 0.3 days, which is coincident with the orbital period. We think that this variability is caused by the satellite. Therefore at least the satellite rotation is synchronous. Whether the rotation is synchronous or double synchronous is not known yet with absolute certainty, but there is considerable evidence that Orcus is spining much faster than 9.5 day. The short-term variability of $\sim$ 0.04 mag and period around 10.5h already reported \citep{Ortiz2006, Thirouin2010} is clear evidence, and there is also a hint for a similar periodicity in the photometry data presented here, although their precision was not sufficiently high. All this would indicate that Orcus primary has not been sufficiently tidally despun to reach a double synchronous state.
If we assume that the initial spin period of Orcus was around its critical value, the total angular momentum lost by the despun to 10 hr would have been gained by the satellite, which would have reached exactly its current configuration if the mass ratio of the system is around 0.09 (the value obtained by assuming that Vanth's albedo is smaller than that of Orcus, which is likely the case according to their very different spectra). This would give support to the idea that the satellite might be the result of a rotational fission.


\begin{acknowledgements}
  We are grateful to P. J. Gutierrez for useful discussions. This research was partially supported by Spanish
grants PCI2005-A7-0180, AYA2008-06202-C03-01,
P07-FQM-02998 and European FEDER funds.
RD acknowledges financial support from the MEC (contract Ram\'on y
Cajal).

\end{acknowledgements}


\longtab{3}{
\begin{longtable}{lcc}
\caption{\label{tbl3} Orcus' satellite E-W positions relative to Orcus (negative to the East) as a function
of date and the RA average residuals for the listed mean julian dates.}\\
\hline\hline Julian Date & E-W Distance (km)  & RA residuals (arcsec) \\
\hline
\endfirsthead
\caption{continued.}\\
\hline\hline Julian Date & E-W Distance (km)  & RA residuals (arcsec) \\
\hline
\endhead
\hline
\endfoot
2455181.78865 & -1979.31 & -0.171 \\
2455183.79607 &  8104.50 & -0.054 \\
2455184.78963 &  8956.02 & -0.046 \\
2455186.77611 &   767.11 &  0.041 \\
2455188.81598 & -8642.56 &  0.181 \\
2455189.82195 & -8496.13 &  0.248 \\
2455190.81837 & -4788.64 & -0.094 \\
2455191.82597 &   946.28 & -0.055 \\
2455192.82017 &  6269.45 & -0.018 \\
2455193.82246 &  8996.14 & -0.068 \\
2455205.79762 &  1106.78 &  0.191 \\
2455206.80230 & -4675.49 & -0.002 \\
2455207.79911 & -8455.05 &  0.171 \\
2455208.78335 & -8722.25 &  0.148 \\
2455209.78888 & -5322.47 &  0.027 \\
2455210.80571 &   387.32 & -0.051 \\
2455211.77190 &  5701.46 & -0.403 \\
2455214.78864 &  4213.96 & -0.036 \\
 \hline\hline
\end{longtable}}

\newpage

\Online

\onlfig{2}{
\begin{figure*}
\centering
\includegraphics[width=16cm]{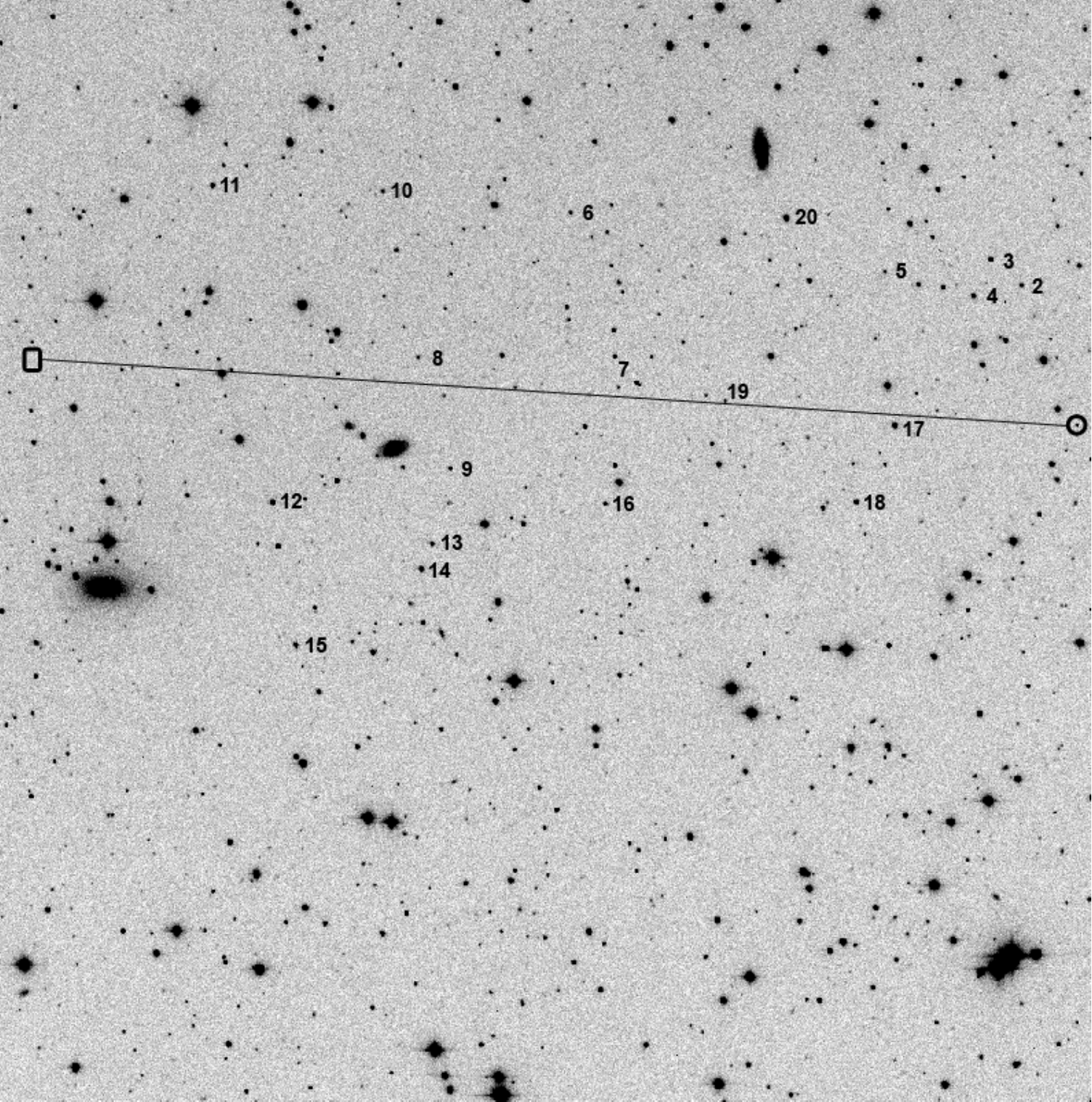}
\caption{Negative image of the central 24.5 $\times$ 24.5 arcmin field that was traversed by Orcus. The position of Orcus at the start of the run is indicated by a circle and its trajectory is shown as a white line. The square indicates where the trajectory ended. North is up, East is to the left. The stars used for the relative photometry analysis are labeled with numbers.}
\label{fig2}
\end{figure*}
}

\onltab{1}{
\longtab{1}{
\begin{longtable}{lcccccccccccc}
\caption{\label{tb1}  Astrometry of the Orcus' system observations, together with the residuals to an orbital fit. The
right ascension and declination are referred to epoch J2000}\\
\hline\hline
 Year & Month & Day &   & RA &   &   & Dec &   &  RA residual  & Dec residual & $\Delta$ & r \\
         &         &   (UT)       &hr & '  & " &$^o$& '   & " &   (arcsec)      &   (arcsec)       &  AU    & AU \\
\hline
\vspace{0.1cm}
\endfirsthead
\caption{continued.}\\
\hline\hline
 Year & Month & Day &   & RA &   &   & Dec &   &  RA residual  & Dec residual & $\Delta$ & r \\
        &             &        &hr& '    & " &$^o$& '  & " &   (arcsec)      &      (arcsec)   &  AU    & AU \\
\hline
\vspace{0.1cm}
\endhead
\hline
\endfoot
2009 & 12 & 16.26831 & 09 & 49 & 13.333 & -06 & 29 & 39.05 & +0.16 & -0.21 & 47.463 & 47.828\\
2009 & 12 & 16.27280 & 09 & 49 & 13.317 & -06 & 29 & 39.19 & +0.05 & -0.28 & 47.463 & 47.828\\
2009 & 12 & 16.27739 & 09 & 49 & 13.310 & -06 & 29 & 39.44 & +0.07 & -0.46 & 47.463 & 47.828\\
2009 & 12 & 16.28198 & 09 & 49 & 13.300 & -06 & 29 & 39.23 & +0.05 & -0.18 & 47.463 & 47.828\\
2009 & 12 & 16.28636 & 09 & 49 & 13.279 & -06 & 29 & 39.20 & -0.14 & -0.08 & 47.463 & 47.828\\
2009 & 12 & 16.29093 & 09 & 49 & 13.270 & -06 & 29 & 39.10 & -0.15 & +0.09 & 47.463 & 47.828\\
2009 & 12 & 16.29543 & 09 & 49 & 13.266 & -06 & 29 & 39.18 & -0.08 & +0.08 & 47.463 & 47.828\\
2009 & 12 & 16.29993 & 09 & 49 & 13.257 & -06 & 29 & 39.31 & -0.09 & +0.02 & 47.463 & 47.828\\
2009 & 12 & 16.30446 & 09 & 49 & 13.248 & -06 & 29 & 39.36 & -0.09 & +0.03 & 47.462 & 47.828\\
2009 & 12 & 16.30895 & 09 & 49 & 13.235 & -06 & 29 & 39.41 & -0.16 & +0.05 & 47.462 & 47.828\\
2009 & 12 & 18.27566 & 09 & 49 & 09.536 & -06 & 30 & 08.05 & +0.10 & +0.05 & 47.433 & 47.828\\
2009 & 12 & 18.28012 & 09 & 49 & 09.546 & -06 & 30 & 08.09 & +0.38 & +0.07 & 47.433 & 47.828\\
2009 & 12 & 18.28466 & 09 & 49 & 09.508 & -06 & 30 & 08.09 & -0.05 & +0.14 & 47.433 & 47.828\\
2009 & 12 & 18.28923 & 09 & 49 & 09.500 & -06 & 30 & 07.97 & -0.03 & +0.32 & 47.432 & 47.828\\
2009 & 12 & 18.29372 & 09 & 49 & 09.494 & -06 & 30 & 08.22 & +0.01 & +0.13 & 47.432 & 47.828\\
2009 & 12 & 18.29830 & 09 & 49 & 09.477 & -06 & 30 & 08.07 & -0.10 & +0.35 & 47.432 & 47.828\\
2009 & 12 & 18.30294 & 09 & 49 & 09.479 & -06 & 30 & 08.22 & +0.07 & +0.26 & 47.432 & 47.828\\
2009 & 12 & 18.30750 & 09 & 49 & 09.474 & -06 & 30 & 08.41 & +0.13 & +0.14 & 47.432 & 47.828\\
2009 & 12 & 18.31199 & 09 & 49 & 09.443 & -06 & 30 & 08.31 & -0.19 & +0.30 & 47.432 & 47.828\\
2009 & 12 & 18.31657 & 09 & 49 & 09.442 & -06 & 30 & 08.64 & -0.07 & +0.03 & 47.432 & 47.828\\
2009 & 12 & 19.26960 & 09 & 49 & 07.542 & -06 & 30 & 21.54 & -0.01 & +0.18 & 47.418 & 47.828\\
2009 & 12 & 19.27375 & 09 & 49 & 07.540 & -06 & 30 & 21.66 & +0.09 & +0.11 & 47.418 & 47.828\\
2009 & 12 & 19.27823 & 09 & 49 & 07.522 & -06 & 30 & 21.49 & -0.04 & +0.34 & 47.418 & 47.828\\
2009 & 12 & 19.28284 & 09 & 49 & 07.521 & -06 & 30 & 21.81 & +0.09 & +0.08 & 47.418 & 47.828\\
2009 & 12 & 19.28733 & 09 & 49 & 07.518 & -06 & 30 & 21.63 & +0.19 & +0.32 & 47.418 & 47.828\\
2009 & 12 & 19.29186 & 09 & 49 & 07.495 & -06 & 30 & 21.86 & -0.01 & +0.16 & 47.417 & 47.828\\
2009 & 12 & 19.29639 & 09 & 49 & 07.492 & -06 & 30 & 22.25 & +0.08 & -0.17 & 47.417 & 47.828\\
2009 & 12 & 19.30100 & 09 & 49 & 07.466 & -06 & 30 & 22.11 & -0.16 & +0.03 & 47.417 & 47.828\\
2009 & 12 & 19.30543 & 09 & 49 & 07.456 & -06 & 30 & 21.97 & -0.17 & +0.23 & 47.417 & 47.828\\
2009 & 12 & 19.30992 & 09 & 49 & 07.458 & -06 & 30 & 21.96 & +0.00 & +0.30 & 47.417 & 47.828\\
2009 & 12 & 21.25594 & 09 & 49 & 03.355 & -06 & 30 & 47.10 & -0.04 & +0.08 & 47.389 & 47.828\\
2009 & 12 & 21.26043 & 09 & 49 & 03.354 & -06 & 30 & 47.07 & +0.09 & +0.17 & 47.388 & 47.828\\
2009 & 12 & 21.26498 & 09 & 49 & 03.346 & -06 & 30 & 46.88 & +0.13 & +0.41 & 47.388 & 47.828\\
2009 & 12 & 21.26954 & 09 & 49 & 03.328 & -06 & 30 & 46.97 & +0.01 & +0.38 & 47.388 & 47.828\\
2009 & 12 & 21.27396 & 09 & 49 & 03.331 & -06 & 30 & 46.88 & +0.20 & +0.52 & 47.388 & 47.828\\
2009 & 12 & 21.27830 & 09 & 49 & 03.301 & -06 & 30 & 47.21 & -0.10 & +0.24 & 47.388 & 47.828\\
2009 & 12 & 21.28287 & 09 & 49 & 03.301 & -06 & 30 & 47.23 & +0.05 & +0.28 & 47.388 & 47.828\\
2009 & 12 & 21.28718 & 09 & 49 & 03.276 & -06 & 30 & 47.59 & -0.18 & -0.03 & 47.388 & 47.828\\
2009 & 12 & 21.29165 & 09 & 49 & 03.286 & -06 & 30 & 47.40 & +0.12 & +0.22 & 47.388 & 47.828\\
2009 & 12 & 21.29621 & 09 & 49 & 03.275 & -06 & 30 & 47.44 & +0.11 & +0.23 & 47.388 & 47.828\\
2009 & 12 & 23.29572 & 09 & 48 & 58.781 & -06 & 31 & 10.72 & +0.22 & +0.17 & 47.359 & 47.828\\
2009 & 12 & 23.30014 & 09 & 48 & 58.762 & -06 & 31 & 10.70 & +0.09 & +0.24 & 47.359 & 47.828\\
2009 & 12 & 23.30474 & 09 & 48 & 58.752 & -06 & 31 & 10.95 & +0.11 & +0.04 & 47.359 & 47.828\\
2009 & 12 & 23.30925 & 09 & 48 & 58.746 & -06 & 31 & 10.94 & +0.18 & +0.10 & 47.359 & 47.828\\
2009 & 12 & 23.31380 & 09 & 48 & 58.731 & -06 & 31 & 10.97 & +0.11 & +0.12 & 47.359 & 47.828\\
2009 & 12 & 23.31820 & 09 & 48 & 58.720 & -06 & 31 & 10.95 & +0.11 & +0.19 & 47.359 & 47.828\\
2009 & 12 & 23.32275 & 09 & 48 & 58.722 & -06 & 31 & 11.13 & +0.30 & +0.06 & 47.359 & 47.828\\
2009 & 12 & 23.32728 & 09 & 48 & 58.697 & -06 & 31 & 11.03 & +0.09 & +0.21 & 47.359 & 47.828\\
2009 & 12 & 23.33167 & 09 & 48 & 58.684 & -06 & 31 & 11.10 & +0.05 & +0.19 & 47.358 & 47.828\\
2009 & 12 & 23.33627 & 09 & 48 & 58.687 & -06 & 31 & 11.11 & +0.26 & +0.23 & 47.358 & 47.828\\
2009 & 12 & 24.30182 & 09 & 48 & 56.393 & -06 & 31 & 21.86 & -0.05 & -0.19 & 47.345 & 47.828\\
2009 & 12 & 24.30607 & 09 & 48 & 56.407 & -06 & 31 & 21.62 & +0.31 & +0.09 & 47.345 & 47.828\\
2009 & 12 & 24.31066 & 09 & 48 & 56.382 & -06 & 31 & 21.61 & +0.11 & +0.15 & 47.345 & 47.828\\
2009 & 12 & 24.31525 & 09 & 48 & 56.362 & -06 & 31 & 21.75 & -0.02 & +0.06 & 47.344 & 47.828\\
2009 & 12 & 24.31958 & 09 & 48 & 56.358 & -06 & 31 & 21.73 & +0.08 & +0.12 & 47.344 & 47.828\\
2009 & 12 & 24.32408 & 09 & 48 & 56.349 & -06 & 31 & 21.79 & +0.11 & +0.11 & 47.344 & 47.828\\
2009 & 12 & 24.32867 & 09 & 48 & 56.352 & -06 & 31 & 21.67 & +0.32 & +0.28 & 47.344 & 47.828\\
2009 & 12 & 24.33315 & 09 & 48 & 56.338 & -06 & 31 & 21.67 & +0.28 & +0.32 & 47.344 & 47.828\\
2009 & 12 & 24.33778 & 09 & 48 & 56.333 & -06 & 31 & 21.78 & +0.37 & +0.26 & 47.344 & 47.828\\
2009 & 12 & 24.34243 & 09 & 48 & 56.312 & -06 & 31 & 21.89 & +0.23 & +0.20 & 47.344 & 47.828\\
2009 & 12 & 25.29835 & 09 & 48 & 53.990 & -06 & 31 & 31.52 & +0.13 & +0.22 & 47.331 & 47.828\\
2009 & 12 & 25.30284 & 09 & 48 & 53.985 & -06 & 31 & 31.89 & +0.22 & -0.10 & 47.331 & 47.828\\
2009 & 12 & 25.30708 & 09 & 48 & 53.936 & -06 & 31 & 31.86 & -0.35 & -0.03 & 47.330 & 47.828\\
2009 & 12 & 25.31158 & 09 & 48 & 53.924 & -06 & 31 & 31.86 & -0.36 & +0.01 & 47.330 & 47.828\\
2009 & 12 & 25.31603 & 09 & 48 & 53.921 & -06 & 31 & 32.11 & -0.24 & -0.19 & 47.330 & 47.828\\
2009 & 12 & 25.32042 & 09 & 48 & 53.922 & -06 & 31 & 32.26 & -0.06 & -0.30 & 47.330 & 47.828\\
2009 & 12 & 25.32499 & 09 & 48 & 53.903 & -06 & 31 & 32.35 & -0.17 & -0.35 & 47.330 & 47.828\\
2009 & 12 & 25.32956 & 09 & 48 & 53.889 & -06 & 31 & 32.36 & -0.20 & -0.31 & 47.330 & 47.828\\
2009 & 12 & 25.33417 & 09 & 48 & 53.878 & -06 & 31 & 32.24 & -0.19 & -0.15 & 47.330 & 47.828\\
2009 & 12 & 25.33866 & 09 & 48 & 53.871 & -06 & 31 & 32.38 & -0.13 & -0.24 & 47.330 & 47.828\\
2009 & 12 & 26.30547 & 09 & 48 & 51.464 & -06 & 31 & 41.61 & -0.13 & -0.29 & 47.317 & 47.828\\
2009 & 12 & 26.31010 & 09 & 48 & 51.449 & -06 & 31 & 41.52 & -0.17 & -0.16 & 47.316 & 47.828\\
2009 & 12 & 26.31459 & 09 & 48 & 51.442 & -06 & 31 & 41.65 & -0.10 & -0.25 & 47.316 & 47.828\\
2009 & 12 & 26.31914 & 09 & 48 & 51.436 & -06 & 31 & 41.78 & -0.02 & -0.34 & 47.316 & 47.828\\
2009 & 12 & 26.32366 & 09 & 48 & 51.418 & -06 & 31 & 41.77 & -0.11 & -0.29 & 47.316 & 47.828\\
2009 & 12 & 26.32818 & 09 & 48 & 51.429 & -06 & 31 & 41.78 & +0.23 & -0.26 & 47.316 & 47.828\\
2009 & 12 & 26.33277 & 09 & 48 & 51.367 & -06 & 31 & 42.10 & -0.52 & -0.53 & 47.316 & 47.828\\
2009 & 12 & 26.33739 & 09 & 48 & 51.391 & -06 & 31 & 41.72 & +0.02 & -0.11 & 47.316 & 47.828\\
2009 & 12 & 26.34193 & 09 & 48 & 51.370 & -06 & 31 & 41.81 & -0.12 & -0.16 & 47.316 & 47.828\\
2009 & 12 & 26.34652 & 09 & 48 & 51.344 & -06 & 31 & 41.97 & -0.33 & -0.28 & 47.316 & 47.828\\
2009 & 12 & 27.29975 & 09 & 48 & 48.920 & -06 & 31 & 50.57 & -0.16 & -0.41 & 47.303 & 47.828\\
2009 & 12 & 27.30432 & 09 & 48 & 48.903 & -06 & 31 & 50.60 & -0.23 & -0.40 & 47.303 & 47.828\\
2009 & 12 & 27.30880 & 09 & 48 & 48.899 & -06 & 31 & 50.55 & -0.11 & -0.31 & 47.303 & 47.828\\
2009 & 12 & 27.31334 & 09 & 48 & 48.891 & -06 & 31 & 50.78 & -0.05 & -0.50 & 47.303 & 47.828\\
2009 & 12 & 27.31786 & 09 & 48 & 48.893 & -06 & 31 & 50.59 & +0.16 & -0.27 & 47.303 & 47.828\\
2009 & 12 & 27.32250 & 09 & 48 & 48.868 & -06 & 31 & 51.12 & -0.03 & -0.76 & 47.303 & 47.828\\
2009 & 12 & 27.32697 & 09 & 48 & 48.852 & -06 & 31 & 50.61 & -0.09 & -0.21 & 47.302 & 47.828\\
2009 & 12 & 27.33153 & 09 & 48 & 48.829 & -06 & 31 & 50.99 & -0.26 & -0.55 & 47.302 & 47.828\\
2009 & 12 & 27.33610 & 09 & 48 & 48.825 & -06 & 31 & 50.74 & -0.13 & -0.26 & 47.302 & 47.828\\
2009 & 12 & 27.34054 & 09 & 48 & 48.816 & -06 & 31 & 51.09 & -0.09 & -0.57 & 47.302 & 47.828\\
2009 & 12 & 28.30215 & 09 & 48 & 46.290 & -06 & 31 & 58.81 & -0.18 & -0.33 & 47.289 & 47.828\\
2009 & 12 & 28.30668 & 09 & 48 & 46.277 & -06 & 31 & 58.80 & -0.19 & -0.28 & 47.289 & 47.828\\
2009 & 12 & 28.31124 & 09 & 48 & 46.263 & -06 & 31 & 58.94 & -0.21 & -0.39 & 47.289 & 47.828\\
2009 & 12 & 28.31578 & 09 & 48 & 46.243 & -06 & 31 & 59.02 & -0.33 & -0.43 & 47.289 & 47.828\\
2009 & 12 & 28.32019 & 09 & 48 & 46.249 & -06 & 31 & 58.84 & -0.06 & -0.22 & 47.289 & 47.828\\
2009 & 12 & 28.32471 & 09 & 48 & 46.228 & -06 & 31 & 58.89 & -0.19 & -0.23 & 47.289 & 47.828\\
2009 & 12 & 28.32923 & 09 & 48 & 46.224 & -06 & 31 & 59.05 & -0.07 & -0.35 & 47.289 & 47.828\\
2009 & 12 & 28.33380 & 09 & 48 & 46.199 & -06 & 31 & 59.00 & -0.25 & -0.27 & 47.289 & 47.828\\
2009 & 12 & 28.33814 & 09 & 48 & 46.196 & -06 & 31 & 59.04 & -0.12 & -0.27 & 47.289 & 47.828\\
2009 & 12 & 28.34268 & 09 & 48 & 46.193 & -06 & 31 & 59.02 & +0.02 & -0.22 & 47.289 & 47.828\\
2010 & 01 & 09.27728 & 09 & 48 & 10.230 & -06 & 32 & 49.95 & +0.47 & +0.39 & 47.140 & 47.828\\
2010 & 01 & 09.28180 & 09 & 48 & 10.226 & -06 & 32 & 50.14 & +0.64 & +0.20 & 47.140 & 47.828\\
2010 & 01 & 09.28628 & 09 & 48 & 10.192 & -06 & 32 & 49.92 & +0.36 & +0.42 & 47.140 & 47.828\\
2010 & 01 & 09.29086 & 09 & 48 & 10.165 & -06 & 32 & 50.28 & +0.19 & +0.06 & 47.140 & 47.828\\
2010 & 01 & 09.29541 & 09 & 48 & 10.151 & -06 & 32 & 50.22 & +0.21 & +0.13 & 47.140 & 47.828\\
2010 & 01 & 09.29985 & 09 & 48 & 10.116 & -06 & 32 & 50.53 & -0.08 & -0.18 & 47.140 & 47.828\\
2010 & 01 & 09.30438 & 09 & 48 & 10.099 & -06 & 32 & 50.28 & -0.11 & +0.07 & 47.140 & 47.828\\
2010 & 01 & 09.30893 & 09 & 48 & 10.083 & -06 & 32 & 50.34 & -0.11 & +0.02 & 47.140 & 47.828\\
2010 & 01 & 09.31342 & 09 & 48 & 10.074 & -06 & 32 & 50.23 & -0.02 & +0.13 & 47.140 & 47.828\\
2010 & 01 & 09.31801 & 09 & 48 & 10.049 & -06 & 32 & 50.24 & -0.16 & +0.12 & 47.140 & 47.828\\
2010 & 01 & 10.28184 & 09 & 48 & 06.814 & -06 & 32 & 50.63 & +0.12 & +0.04 & 47.129 & 47.828\\
2010 & 01 & 10.28644 & 09 & 48 & 06.789 & -06 & 32 & 51.18 & -0.01 & -0.51 & 47.129 & 47.828\\
2010 & 01 & 10.29102 & 09 & 48 & 06.775 & -06 & 32 & 50.58 & +0.02 & +0.09 & 47.129 & 47.828\\
2010 & 01 & 10.29558 & 09 & 48 & 06.758 & -06 & 32 & 50.72 & +0.00 & -0.05 & 47.129 & 47.828\\
2010 & 01 & 10.30008 & 09 & 48 & 06.745 & -06 & 32 & 50.34 & +0.04 & +0.33 & 47.129 & 47.828\\
2010 & 01 & 10.30462 & 09 & 48 & 06.723 & -06 & 32 & 50.55 & -0.05 & +0.12 & 47.129 & 47.828\\
2010 & 01 & 10.30909 & 09 & 48 & 06.706 & -06 & 32 & 50.48 & -0.08 & +0.19 & 47.129 & 47.828\\
2010 & 01 & 10.31362 & 09 & 48 & 06.684 & -06 & 32 & 50.47 & -0.17 & +0.20 & 47.129 & 47.828\\
2010 & 01 & 10.31815 & 09 & 48 & 06.677 & -06 & 32 & 50.58 & -0.04 & +0.09 & 47.129 & 47.828\\
2010 & 01 & 10.32258 & 09 & 48 & 06.658 & -06 & 32 & 50.56 & -0.09 & +0.11 & 47.128 & 47.828\\
2010 & 01 & 11.27892 & 09 & 48 & 03.401 & -06 & 32 & 50.26 & +0.19 & +0.13 & 47.118 & 47.828\\
2010 & 01 & 11.28336 & 09 & 48 & 03.384 & -06 & 32 & 50.19 & +0.17 & +0.20 & 47.118 & 47.828\\
2010 & 01 & 11.28785 & 09 & 48 & 03.361 & -06 & 32 & 50.10 & +0.07 & +0.29 & 47.118 & 47.828\\
2010 & 01 & 11.29245 & 09 & 48 & 03.349 & -06 & 32 & 50.04 & +0.13 & +0.34 & 47.118 & 47.828\\
2010 & 01 & 11.29683 & 09 & 48 & 03.329 & -06 & 32 & 50.17 & +0.06 & +0.21 & 47.118 & 47.828\\
2010 & 01 & 11.30130 & 09 & 48 & 03.303 & -06 & 32 & 50.39 & -0.09 & -0.01 & 47.118 & 47.828\\
2010 & 01 & 11.30579 & 09 & 48 & 03.307 & -06 & 32 & 50.17 & +0.20 & +0.21 & 47.118 & 47.828\\
2010 & 01 & 11.31043 & 09 & 48 & 03.295 & -06 & 32 & 50.50 & +0.27 & -0.13 & 47.118 & 47.828\\
2010 & 01 & 11.31486 & 09 & 48 & 03.277 & -06 & 32 & 50.11 & +0.23 & +0.26 & 47.118 & 47.828\\
2010 & 01 & 11.31930 & 09 & 48 & 03.263 & -06 & 32 & 50.17 & +0.26 & +0.20 & 47.118 & 47.828\\
2010 & 01 & 12.26323 & 09 & 47 & 59.959 & -06 & 32 & 49.52 & -0.12 & -0.01 & 47.107 & 47.828\\
2010 & 01 & 12.26767 & 09 & 47 & 59.964 & -06 & 32 & 49.48 & +0.19 & +0.03 & 47.107 & 47.828\\
2010 & 01 & 12.27218 & 09 & 47 & 59.939 & -06 & 32 & 49.50 & +0.06 & +0.00 & 47.107 & 47.828\\
2010 & 01 & 12.27659 & 09 & 47 & 59.936 & -06 & 32 & 49.60 & +0.25 & -0.10 & 47.107 & 47.828\\
2010 & 01 & 12.28107 & 09 & 47 & 59.918 & -06 & 32 & 49.63 & +0.22 & -0.14 & 47.107 & 47.828\\
2010 & 01 & 12.28548 & 09 & 47 & 59.914 & -06 & 32 & 49.48 & +0.40 & +0.01 & 47.107 & 47.828\\
2010 & 01 & 12.29011 & 09 & 47 & 59.886 & -06 & 32 & 49.26 & +0.22 & +0.22 & 47.107 & 47.828\\
2010 & 01 & 12.29460 & 09 & 47 & 59.853 & -06 & 32 & 49.43 & -0.03 & +0.05 & 47.107 & 47.828\\
2010 & 01 & 12.29904 & 09 & 47 & 59.852 & -06 & 32 & 49.09 & +0.19 & +0.38 & 47.107 & 47.828\\
2010 & 01 & 12.30352 & 09 & 47 & 59.824 & -06 & 32 & 49.58 & +0.01 & -0.11 & 47.107 & 47.828\\
2010 & 01 & 13.26833 & 09 & 47 & 56.432 & -06 & 32 & 48.00 & +0.12 & +0.00 & 47.097 & 47.828\\
2010 & 01 & 13.27293 & 09 & 47 & 56.405 & -06 & 32 & 48.06 & -0.04 & -0.07 & 47.097 & 47.828\\
2010 & 01 & 13.27753 & 09 & 47 & 56.397 & -06 & 32 & 48.14 & +0.09 & -0.15 & 47.097 & 47.828\\
2010 & 01 & 13.28210 & 09 & 47 & 56.353 & -06 & 32 & 47.80 & -0.32 & +0.18 & 47.097 & 47.828\\
2010 & 01 & 13.28664 & 09 & 47 & 56.373 & -06 & 32 & 47.93 & +0.23 & +0.04 & 47.097 & 47.828\\
2010 & 01 & 13.29116 & 09 & 47 & 56.366 & -06 & 32 & 47.93 & +0.36 & +0.03 & 47.097 & 47.828\\
2010 & 01 & 13.29566 & 09 & 47 & 56.337 & -06 & 32 & 47.80 & +0.17 & +0.15 & 47.097 & 47.828\\
2010 & 01 & 13.30027 & 09 & 47 & 56.308 & -06 & 32 & 47.60 & -0.01 & +0.35 & 47.097 & 47.828\\
2010 & 01 & 13.30488 & 09 & 47 & 56.293 & -06 & 32 & 48.00 & +0.02 & -0.06 & 47.097 & 47.828\\
2010 & 01 & 13.30932 & 09 & 47 & 56.263 & -06 & 32 & 48.03 & -0.19 & -0.10 & 47.097 & 47.828\\
2010 & 01 & 14.28568 & 09 & 47 & 52.794 & -06 & 32 & 45.95 & +0.07 & -0.11 & 47.087 & 47.828\\
2010 & 01 & 14.29017 & 09 & 47 & 52.786 & -06 & 32 & 45.76 & +0.20 & +0.07 & 47.086 & 47.828\\
2010 & 01 & 14.29460 & 09 & 47 & 52.760 & -06 & 32 & 45.59 & +0.05 & +0.23 & 47.086 & 47.828\\
2010 & 01 & 14.29903 & 09 & 47 & 52.756 & -06 & 32 & 45.63 & +0.23 & +0.18 & 47.086 & 47.828\\
2010 & 01 & 14.30353 & 09 & 47 & 52.703 & -06 & 32 & 45.75 & -0.31 & +0.05 & 47.086 & 47.828\\
2010 & 01 & 14.30790 & 09 & 47 & 52.736 & -06 & 32 & 45.37 & +0.42 & +0.42 & 47.086 & 47.828\\
2010 & 01 & 14.31242 & 09 & 47 & 52.680 & -06 & 32 & 45.70 & -0.17 & +0.08 & 47.086 & 47.828\\
2010 & 01 & 14.31687 & 09 & 47 & 52.668 & -06 & 32 & 45.99 & -0.11 & -0.22 & 47.086 & 47.828\\
2010 & 01 & 14.32121 & 09 & 47 & 52.642 & -06 & 32 & 45.96 & -0.26 & -0.21 & 47.086 & 47.828\\
2010 & 01 & 14.32564 & 09 & 47 & 52.623 & -06 & 32 & 45.57 & -0.30 & +0.17 & 47.086 & 47.828\\
2010 & 01 & 15.25154 & 09 & 47 & 49.283 & -06 & 32 & 43.37 & -0.23 & -0.17 & 47.077 & 47.828\\
2010 & 01 & 15.25608 & 09 & 47 & 49.252 & -06 & 32 & 43.00 & -0.44 & +0.19 & 47.077 & 47.828\\
2010 & 01 & 15.26066 & 09 & 47 & 49.243 & -06 & 32 & 43.23 & -0.32 & -0.05 & 47.077 & 47.828\\
2010 & 01 & 15.26521 & 09 & 47 & 49.224 & -06 & 32 & 43.16 & -0.36 & +0.00 & 47.077 & 47.828\\
2010 & 01 & 15.26968 & 09 & 47 & 49.192 & -06 & 32 & 43.33 & -0.59 & -0.18 & 47.077 & 47.828\\
2010 & 01 & 15.27425 & 09 & 47 & 49.193 & -06 & 32 & 43.05 & -0.32 & +0.09 & 47.077 & 47.828\\
2010 & 01 & 15.27873 & 09 & 47 & 49.171 & -06 & 32 & 43.03 & -0.40 & +0.09 & 47.077 & 47.828\\
2010 & 01 & 15.28308 & 09 & 47 & 49.160 & -06 & 32 & 43.09 & -0.32 & +0.02 & 47.077 & 47.828\\
2010 & 01 & 15.28765 & 09 & 47 & 49.149 & -06 & 32 & 43.12 & -0.23 & -0.03 & 47.076 & 47.828\\
2010 & 01 & 15.29215 & 09 & 47 & 49.131 & -06 & 32 & 43.06 & -0.25 & +0.02 & 47.076 & 47.828\\
2010 & 01 & 18.27056 & 09 & 47 & 38.125 & -06 & 32 & 31.26 & +0.06 & +0.05 & 47.048 & 47.828\\
2010 & 01 & 18.27514 & 09 & 47 & 38.103 & -06 & 32 & 31.21 & +0.00 & +0.07 & 47.048 & 47.828\\
2010 & 01 & 18.27968 & 09 & 47 & 38.069 & -06 & 32 & 31.32 & -0.25 & -0.06 & 47.048 & 47.828\\
2010 & 01 & 18.28411 & 09 & 47 & 38.080 & -06 & 32 & 31.48 & +0.17 & -0.24 & 47.048 & 47.828\\
2010 & 01 & 18.28857 & 09 & 47 & 38.063 & -06 & 32 & 31.24 & +0.17 & -0.02 & 47.048 & 47.828\\
2010 & 01 & 18.29304 & 09 & 47 & 38.044 & -06 & 32 & 31.86 & +0.14 & -0.66 & 47.048 & 47.828\\
2010 & 01 & 18.29760 & 09 & 47 & 38.037 & -06 & 32 & 31.43 & +0.29 & -0.26 & 47.048 & 47.828\\
2010 & 01 & 18.30224 & 09 & 47 & 38.006 & -06 & 32 & 31.33 & +0.10 & -0.18 & 47.048 & 47.828\\
2010 & 01 & 18.30683 & 09 & 47 & 37.993 & -06 & 32 & 31.15 & +0.17 & -0.02 & 47.048 & 47.828\\
2010 & 01 & 18.31138 & 09 & 47 & 37.962 & -06 & 32 & 31.47 & -0.04 & -0.36 & 47.048 & 47.828\\
\end{longtable}
}}

\onltab{2}{
\begin{table*}
\caption{ Orcus'system relative photometry. Date is expressed as reduced Julian Date (JD-2450000).}
\label{tbl2}
\begin{tabular}{lccccccc}
\hline \hline \vspace{0.1cm}
Reduced & Relative  & Reduced & Relative  & Reduced & Relative & Reduced & Relative \\
JD      & magnitude &  JD     & magnitude &   JD    & magnitude &   JD    & magnitude \\
 \hline
5181.76657 &  0.0700 & 5186.78113 &  0.0159 & 5191.82644 & -0.0220 & 5207.81312 & -0.0483 \\
5181.77106 & -0.0468 & 5186.78991 & -0.0345 & 5191.83103 &  0.0428 & 5207.81756 & -0.0851 \\
5181.77565 & -0.0293 & 5186.79447 & -0.0125 & 5191.83565 & -0.0013 & 5208.76149 & -0.0352 \\
5181.78024 &  0.0591 & 5188.79398 & -0.1081 & 5193.80041 &  0.1017 & 5208.76593 & -0.0291 \\
5181.78919 & -0.0379 & 5188.79840 & -0.0735 & 5193.80494 &  0.0185 & 5208.77044 & -0.0158 \\
5181.79819 &  0.0229 & 5188.80300 & -0.0651 & 5193.80950 &  0.0439 & 5208.77485 & -0.0870 \\
5181.80272 & -0.1085 & 5188.80751 &  0.0023 & 5193.81845 & -0.0190 & 5208.77933 &  0.0290 \\
5181.80721 & -0.0645 & 5188.81206 &  0.0348 & 5193.82297 &  0.0333 & 5208.78374 &  0.0251 \\
5183.77838 & -0.0051 & 5188.81646 & -0.0522 & 5193.82749 & -0.0251 & 5208.79286 & -0.0355 \\
5183.78292 &  0.0446 & 5188.82101 & -0.0715 & 5193.83640 &  0.0875 & 5208.79730 &  0.0099 \\
5183.78749 &  0.0059 & 5188.82993 &  0.0119 & 5193.84094 & -0.0120 & 5208.80178 & -0.0113 \\
5183.79198 &  0.0807 & 5188.83453 & -0.0975 & 5205.78006 &  0.0433 & 5209.76659 & -0.0659 \\
5183.79656 &  0.0579 & 5189.80008 & -0.0183 & 5205.78454 &  0.0435 & 5209.78036 & -0.0650 \\
5183.80120 &  0.0530 & 5189.80433 & -0.0351 & 5205.78912 &  0.0616 & 5209.78490 & -0.0832 \\
5183.80576 &  0.0544 & 5189.80892 & -0.0073 & 5205.79367 &  0.0674 & 5209.78942 &  0.0341 \\
5183.81025 &  0.0114 & 5189.81351 & -0.0589 & 5205.79811 &  0.0236 & 5209.79392 &  0.0133 \\
5183.81483 &  0.0365 & 5189.81784 & -0.0888 & 5205.80719 & -0.0184 & 5211.74980 &  0.0038 \\
5184.76786 &  0.0271 & 5189.82234 &  0.0119 & 5205.81168 &  0.0858 & 5211.75434 &  0.0100 \\
5184.77201 & -0.0372 & 5189.82693 & -0.0097 & 5205.81627 &  0.1208 & 5211.75892 & -0.0305 \\
5184.77649 &  0.0685 & 5189.83141 & -0.0042 & 5206.78010 & -0.0273 & 5211.76347 &  0.0500 \\
5184.78110 &  0.0190 & 5189.83604 & -0.0046 & 5206.78470 &  0.0605 & 5211.76794 &  0.0271 \\
5184.78559 &  0.0129 & 5190.80110 & -0.0080 & 5206.78928 & -0.0156 & 5211.77251 &  0.1004 \\
5184.79012 & -0.0079 & 5190.80534 & -0.0948 & 5206.79384 &  0.0257 & 5211.77699 & -0.0215 \\
5184.79465 &  0.0180 & 5190.81429 &  0.0750 & 5206.79834 &  0.1009 & 5211.78591 &  0.0324 \\
5184.79926 &  0.0518 & 5190.81868 & -0.0245 & 5206.80735 &  0.0812 & 5211.79041 & -0.0203 \\
5184.80369 &  0.0131 & 5190.83243 & -0.0444 & 5206.81188 & -0.0168 & 5214.76882 & -0.0424 \\
5184.80818 & -0.0374 & 5190.83692 &  0.0088 & 5206.82084 & -0.0564 & 5214.77340 &  0.0396 \\
5186.75420 &  0.0194 & 5191.80373 & -0.0117 & 5207.77718 & -0.0695 & 5214.77794 & -0.0597 \\
5186.75869 & -0.0386 & 5191.81285 & -0.0102 & 5207.78162 &  0.0488 & 5214.79586 &  0.0273 \\
5186.76324 &  0.0627 & 5191.81740 &  0.0351 & 5207.79071 & -0.0655 & 5214.80050 & -0.0031 \\
5186.76780 &  0.0527 & 5191.82192 & -0.0444 & 5207.79509 & -0.0042 & 5214.80509 & -0.0454 \\
\hline
\end{tabular}
\end{table*}
}


\end{document}